\documentclass[preprint,12pt]{elsarticle}
\usepackage{natbib}
\usepackage{graphicx, subfigure, float}
\usepackage{amsmath,amssymb}
\usepackage{epstopdf}
\usepackage{linguex}

\journal{Proc R Soc A\ }

\newcommand{\pref}[1]{(\ref{#1})}

\newcommand{\Eq}[1]{Eq.~(\ref{#1})}
\newcommand{\Fig}[1]{Fig.~\ref{#1}}

\renewcommand{\vec}[1]{\mbox{\boldmath$#1$}}

\begin{document}

\title{Detecting event-related recurrences by symbolic analysis: Applications to human language processing}

\author{Peter beim Graben\corref{cor1}}
\address{Bernstein Center for Computational Neuroscience Berlin, Germany}
\address{Department of German Studies and Linguistics,
 Humboldt-Universit\"at zu Berlin}
\ead{peter.beim.graben@hu-berlin.de}
\ead[url]{www.beimgraben.info}
\cortext[cor1]{Department of German Language and Linguistics \\
 Humboldt-Universit\"at zu Berlin \\
 Unter den Linden 6 \\
 D -- 10099 Berlin \\
 Phone: +49-30-2093-9632 \\
 Fax: +49-30-2093-9729
 }

\author{Axel Hutt\corref{nc}}
\address{Team Neurosys, INRIA CR Nancy, France}

\begin{abstract}
Quasistationarity is ubiquitous in complex dynamical systems. In brain dynamics there is ample evidence that event-related potentials reflect such quasistationary states. In order to detect them from time series, several segmentation techniques have been proposed. In this study we elaborate a recent approach for detecting quasistationary states as recurrence domains by means of recurrence analysis and subsequent symbolisation methods. As a result, recurrence domains are obtained as partition cells that can be further aligned and unified for different realisations. We address two pertinent problems of contemporary recurrence analysis and present possible solutions for them.
\end{abstract}

\maketitle

\section{Introduction}

The brain is an open complex system that receives a broad range of sensational physiological inputs, performs mental tasks and triggers various sorts of physiological output. The corresponding neural information is processed in time either sequentially, e.g., as in low-level response to sensory input~\cite{Naatanen+Picton87, Regan, Basar80, Woods90}, or parallel in different brain areas as in most higher level neural processing steps, e.g., in motor action~\cite{Latash_etal07} and cognition~\cite{Ungerleider+Pessoa08, Haxby_etal00}. In both sequential and parallel processing steps, electrophysiological recordings exhibit a sequence of quasistationary signal states reflecting the underlying neural processing steps~\cite{WackermannEA93, YildizKiebel2011, RabinovichHuertaLaurent08, Gazzaniga_etal13}. Here quasistationary states evolve on a large time scale while transitions between them are much faster.

Sequential patterns of quasistationary states have been observed in various quite different natural systems. For instance, the chaotic Lorenz attractor exhibits two quasistationary states (the so-called {\em Lorenz wings}) which are visited recurrently over time. The Lorenz model is one of the first nonlinear atmospheric models applied in weather forecast~\cite{Lorenz63} while it also describes the dynamics in solid-state laser systems~\cite{Haken}. In neural systems, recurrent temporal activity patterns are known from the replay of memory in hippocampus~\cite{Pavlidis+Winson}, in bird songs~\cite{YildizKiebel2011} and in electroencephalographic data (EEG) obtained during epileptic seizures~\cite{Uhl96, Wendling, AllefeldAtmanspacherWackermann09}. Some more detailed analysis of the dynamics of quasistationary neural states reveal properties of metastable attractors, which attract and repel in different directions in phase space. Good examples for such metastable attractors are saddle nodes with an attracting and a repelling manifold. Hence a sequence of underlying metastable attractors in neural activity generates measurement signals which exhibit a sequence of quasistationary states. Mazor and Laurent~\cite{MazorLaurent05} found such a sequence of metastable attractors in the activity of projection neurons in the insect olfactory bulb~\cite{MazorLaurent05, RabinovichHuertaLaurent08}, while Freeman and Rogers~\cite{Freeman+Rogers02} revealed sequential rapid phase transitions between spatially synchronised states in EEG signals. Most recently, Hudson et al.~\cite{HudsonEA14} discovered metastable transition networks in the recovery from anesthesia.

During cognitive tasks, Lehmann and colleagues~\cite{Lehmann80,LehmannOzakiPal87, WackermannEA93, Brandeis, LehmannPascualMichel09} observed segments of quasistationary EEG topographies, which they called \emph{brain microstates}. The EEG related to cognitive tasks exhibits a sequence of signal components and is termed {\em event-related potential} (ERP). Lehmann et al. were able to identify the extracted microstates with ERP-components which in turn are signal components well-established in neuropsychology as markers of certain neural cognitive processes. This link between quasistationary signal states and ERP-components corroborates the working hypotheses of the present study: ERP components reflect quasistable attractor states in brain dynamics~\cite{Hutt+Munk06, HuttRiedel03, Hutt+Schrauf07, Hutt04}. This hypothesis indicates that the brain performs information processing in a sequence of steps \cite{GrabenPotthast09a} that is reflected in quasistationary signal states. Therefore, improving the detection of such states in neurophysiological data promises to improve the study of the corresponding states, their neural origin and essentially their meaning in neural information processing.

For the detection of quasistationary states from real-world data several methods have been proposed. Obviously, one may first plot the signal trajectory in signal space and extract the time windows of apparent quasistationary by visual inspection. To this end, the dimensionality of multivariate signals may be reduced by principle component analysis (PCA) or related techniques, cf.~\cite{MazorLaurent05, HudsonEA14}. These methods give first insights into the dynamics, but do allow neither classification of the quasistationary states nor extraction of their durations and their transition phases. More advanced methods for segmenting quasistationary states have to consider (\emph{i}) nonstationary signal features, i.e., applying temporally instantaneous methods, (\emph{ii}) multivariate time series, i.e., multidimensional signals exhibiting heterogeneous signal characteristics in different dimensions, and (\emph{iii}) low signal-to-noise ratio where the noise level is not known {\em a priori}.

Techniques applied in the context of brain computer interfaces aim to extract signal features from multivariate signals~\cite{Bostanov04,Liang+Bougrain12}. Most of them perform a first supervised learning task in order to determine the specific features to be extracted from the EEG. However this learning phase is not applicable in the detection of quasistationary states. The present work aims to extract signal features that are unknown {\em a priori} and which may differ slightly between experimental runs.

One of the first successful methods dealing with these signal characteristics is the microstate segmentation technique of Lehmann et al.~\cite{Lehmann80, WackermannEA93, Brandeis, Pascual-MarquiMichelLehmann95, LehmannPascualMichel09}, which extracts transient quasistationary states in multivariate EEG signals based on the similarity of their spatial scalp distributions. It considers the multivariate EEG as a temporal sequence of spatial activity maps and extracts the time windows of the microstates by computing the temporal difference between successive maps. This procedure allows to compute the time windows of microstates from the signal and classifies them by the spatial averages over EEG electrodes in the extracted state time window. This method works well for high-dimensional EEG signals with good spatial resolution and high signal-to-noise ratio. Other similar methods extract signal features by clustering techniques~\cite{HuttRiedel03, Hutt+Schrauf07, Hutt04} taking into account the similarity of the time series in different electrodes. Such methods are less sensitive to noise compared to the microstates method since they avoid the computation of temporal differences.

Although methods for single experimental EEG data sets exist, they imply a missing statistical evaluation of the detected time windows, i.e., one can not ensure that the states found reflect an underlying neural metastable attractor and not just noise. Hence, to evaluate statistically the gained results, it is necessary to perform a statistical evaluation study over several trials. However, in turn, this implies to map detected temporal segments of different trials to each other and, in the best case, align them to each other. This task is difficult to perform since single trials are known to exhibit temporal jitter to each other~\cite{QuianQuiroga, OuyangHerzmannEA11}, i.e., they are delayed and shortened to each other. Previous studies have attempted successfully to deal with this problem by applying dynamic time warping on EEG~\cite{Huang+Jansen85}. This technique considers a certain signal template and detects it in the data set under study with respect to dilation and shrinking operations. The method is not well adjusted to our problem since it is based on a pre-defined signal template which is not known in practice.

The present work elaborates our recently developed recurrence domain analysis method for further improving the temporal segmentation of multivariate signals into quasistationary states by means of symbolic dynamics~\cite{GrabenHutt13}. This method further relaxes the similarity assumption in single time series in the data set under study and is less sensitive to noise. Moreover, we demonstrate how to compute an optimal symbolic sequence in each trial by aligning sequences of quasistationary states that are extracted in each trial by recurrence domain analysis. We introduce a Hausdorff partition clustering method that performs this alignment. The subsequent section introduces to the technique of recurrence domain analysis and applies a new method on single realisations, i.e. single data sets. Then, several extensions of this method are presented followed by a novel method based on the Hausdorff distance to align multiple symbolic sequences gained from multiple data sets.

\section{Symbolic Recurrence Domain Analysis}
\label{sec:srda}

Nonlinear dynamical systems generally obey Poincar\'{e}'s famous \emph{recurrence theorem}\footnote{
    Under the assumptions that the dynamics possesses an invariant measure and is restricted to a finite portion of phase space.
}
\cite{Poincare1890} which states that almost all trajectories starting in a ``ball'' $B_\varepsilon(\vec{x}_0)$ of radius $\varepsilon > 0$ centered at an initial condition $\vec{x}_0 \in X$ (where $X \subset \mathbb{R}^d$ denotes the system's phase space of dimension $d$) return infinitely often to $B_\varepsilon(\vec{x}_0)$ as time elapses. For time-discrete (or discretely sampled) dynamical systems, these recurrences can be visualized by means of Eckmann et al.'s \cite{EckmannOliffsonRuelle87} recurrence plot (RP) technique where the element
\begin{equation}
\label{eq:rp}
    R_{ij} = \begin{cases}
    1 & \text{ if  }  || \vec{x}_i - \vec{x}_j || < \varepsilon \\
    0 & \text{ else }
    \end{cases}
\end{equation}
of the \emph{recurrence matrix} $\vec{R} = (R_{ij})$ is unity if the state $\vec{x}_j$ at time $j$ belongs to an $\varepsilon$-ball centered at state $\vec{x}_i$ at time $i$, i.e.,  $\vec{x}_j \in B_\varepsilon(\vec{x}_i)$, and zero otherwise \cite{EckmannOliffsonRuelle87, MarwanKurths05}.

Often, RPs exhibit a characteristic ``checkerboard texture'' \cite{EckmannOliffsonRuelle87} indicating the system's \emph{recurrence domains} which are quasistable regions in phase space with relatively large dwell that are connected by transients. Every time the systems returns into one of these recurrence domains, the RP displays such a texture, that is symmetric with respect to parallel translations along the horizontal and vertical time axes. Paradigmatic examples for recurrence domains are, e.g., the wings of the Lorenz attractor centered around its unstable foci \cite{Lorenz63}, or saddle sets (such as saddle nodes or saddle tori) that are connected by \emph{stable heteroclinic sequences (SHS)} \cite{RabinovichHuertaLaurent08, AfraimovichRabinovichVarona04, AfraimovichZhigulinRabinovich04}.

In a recent study \cite{GrabenHutt13}, we have proposed a method for the detection of recurrence domains by means of symbolic dynamics \cite{DawFinneyTracy03, Hao91, LindMarcus95}. Our approach is motivated by the fact that a recurrence $R_{ij} = 1$ leads to overlapping $\varepsilon$-balls $B_\varepsilon(\vec{x}_i) \cap B_\varepsilon(\vec{x}_j) \ne \emptyset$ that can be merged together into equivalence classes which eventually partition, together with their complements and the still isolated balls,  the phase space into its respective recurrence domains. Let $\mathcal{P} = \{ A_k \subset X | 1 \le k \le n \}$ be a finite partition of the phase space $X$ into $n$ disjoint sets $A_k$. Then, the discrete trajectory $(\vec{x}_i)$ is mapped onto a symbolic sequence $s = (s_i)$ according to the cells of $\mathcal{P}$ being visited by the states $\vec{x}_i$, i.e., $s_i = k$ when $\vec{x}_i \in A_k$.

In \cite{GrabenHutt13} we suggested to create the symbolic sequence $s = (s_i)$ from an initial sequence of time indices $r_i = i$ subjected to a rewriting grammar \cite{HopcroftUllman79} to which we refer henceforth as to the \emph{recurrence grammar} since this grammar is simply another interpretation of the recurrence matrix $\vec{R}$ obtained in \Eq{eq:rp}. It contains \emph{rewriting rules} that replace large time indices $i$ by smaller ones $j$ ($i > j$) when the corresponding states $\vec{x}_i, \vec{x}_j$ are recurrent, $R_{ij} = 1$. Moreover, if three states $\vec{x}_i, \vec{x}_j, \vec{x}_k$ are recurrent, $R_{ij} = 1$ and $R_{ik} = 1$, for $i > j > k$, the grammar replaces the two larger indices $i, j$ by the smallest one $k$. In other words, the recurrence grammar comprises rewriting rules
\begin{eqnarray}
\label{eq:recgram1}
&&
\begin{aligned}
    i &\to j
\end{aligned}
\quad \text{if } i > j \text{ and } R_{ij} = 1 \\
\label{eq:recgram2}
&&\left.
\begin{aligned}
    i &\to k \\
    j &\to k
\end{aligned}
\right\}
\quad \text{if } i > j > k \text{ and }  R_{ij} = 1,  R_{ik} = 1 \:.
\end{eqnarray}

We illustrate the action of the recurrence grammar with a simple example. Let us assume that a trajectory consists of only five data points $(\vec{x}_1, \vec{x}_2, \dots, \vec{x}_5)$ giving rise to the recurrence matrix
\begin{equation}
\label{eq:examp}
    \vec{R} = \begin{pmatrix}
                1 & 1 & 0 & 0 & 1 \\
                1 & 1 & 0 & 1 & 0 \\
                0 & 0 & 1 & 1 & 0 \\
                0 & 1 & 1 & 1 & 0 \\
                1 & 0 & 0 & 0 & 1 \\
              \end{pmatrix} \:.
\end{equation}
The algorithm commences with the 5th row, detecting a recurrence $R_{51} = 1$. Since $5>1$, a rewriting rule $5 \to 1$ is generated. Because the next recurrence in row 5 is trivial, the algorithm continues with row 4, where $R_{42} = R_{43} = 1$. Now, two rules $4 \to 2$ and $3 \to 2$ are created. The remaining rows 3, 2 and 1 can be neglected due to the symmetry. Thus, we obtain the following recurrence grammar
\begin{align}
\label{eq:exgram}
    5 & \to 1 \nonumber \\
    4 & \to 2 \\
    3 & \to 2 \nonumber \:.
\end{align}
Recursively applying this grammar to the series of time indices $r = 12345$ yields $s = 12221$, i.e. a system with two recurrence domains 1 and 2. Note, that recurrence grammars could be ambiguous, i.e. one left-hand symbol could be expanded into different right-hand symbols by several rules. Yet, this is not a problem for the segmentation of the trajectory into recurrence domains when the grammar is recursively applied. After at most two iterations all ambiguities become resolved through the smallest time index symbolising the earliest recurrence domain.

Furthermore, the symbolic alphabet of the rewritten sequence may contain several gaps. We therefore augmented our algorithm by a search and replace mechanism recoding unused smaller indices by larger ones, thereby squeezing the symbolic repertoire together.

\subsection{Numerical Simulations}
\label{sec:simul}

For further illustration and validation of our algorithm, we carry out a numerical simulation study of a stable heteroclinic contour, i.e. a closed stable heteroclinic sequence between three saddle nodes \cite{RabinovichHuertaLaurent08, AfraimovichRabinovichVarona04, AfraimovichZhigulinRabinovich04} that may be regarded as a model for event-related brain potentials, according to our working hypothesis \cite{HuttRiedel03, Hutt04}.

The stable heteroclinic contour was obtained from a generalized Lotka-Volterra dynamics of $n = 3$ competing populations
\begin{eqnarray}
\label{eq:lovolt}
   \frac{\mathrm{d} x_k(t)}{\mathrm{d} t} = x_k \left( \sigma_k  - \sum_{j=1}^n \rho_{kj} x_j \right)
\end{eqnarray}
with growth rates $\sigma_k >0$, and interaction weights $\rho_{kj} >0$, $\rho_{kk} = 1$. The particular parameter settings were $\sigma_1 = 1$, $\sigma_2 = 1.2$, and $\sigma_3 = 1.6$.

Lotka-Volterra systems are well-studied in theoretical ecology where they describe the intra- and inter-specific interactions, i.e. competition for limited resources and predator-prey networks, among $n$ populations with abundances $x_k$ in an ecosystem. Moreover, these systems have a longstanding tradition in computational neuroscience \cite{Cowan14, RabinovichEA06} as a suitable  approximation \cite{FukaiTanaka97} of the Wilson-Cowan neural mass model \cite{WilsonCowan72}.

The Lotka-Volterra equations often describe limit cycles in appropriate parameter regimes. However, they also exhibit stable heteroclinic sequences (SHS) \cite{RabinovichHuertaLaurent08, AfraimovichRabinovichVarona04, AfraimovichZhigulinRabinovich04} and stable heteroclinic channels (SHC) \cite{RabinovichEA08, RabinovichHuertaLaurent08} for asymmetric interactions which is of particular importance for neural applications.

\subsubsection{Recurrence Grammars}
\label{sec:rg}

\begin{figure}[H]
\centering
\subfigure[]{\includegraphics[width=0.45\textwidth]{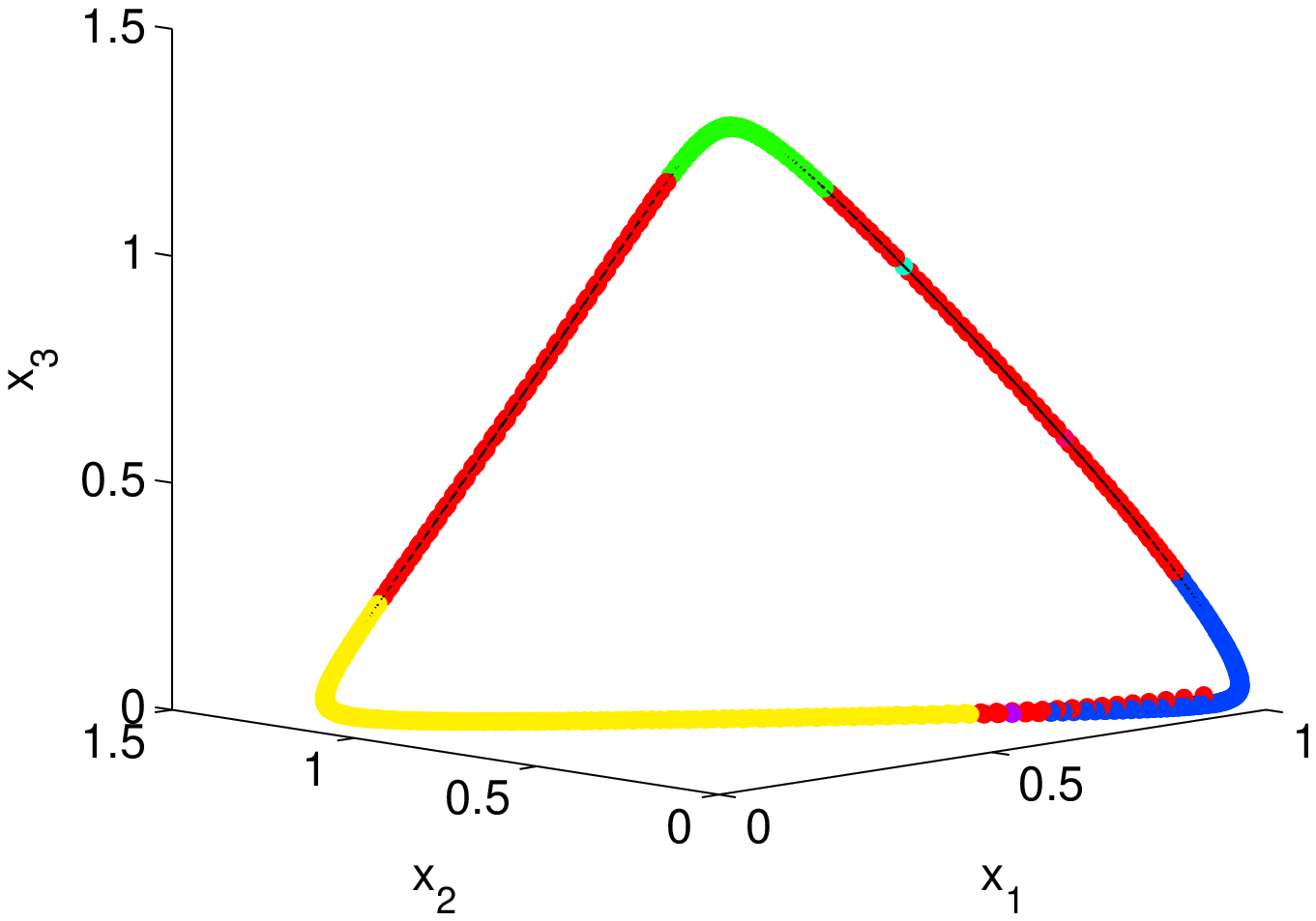}}
\subfigure[]{\includegraphics[width=0.45\textwidth]{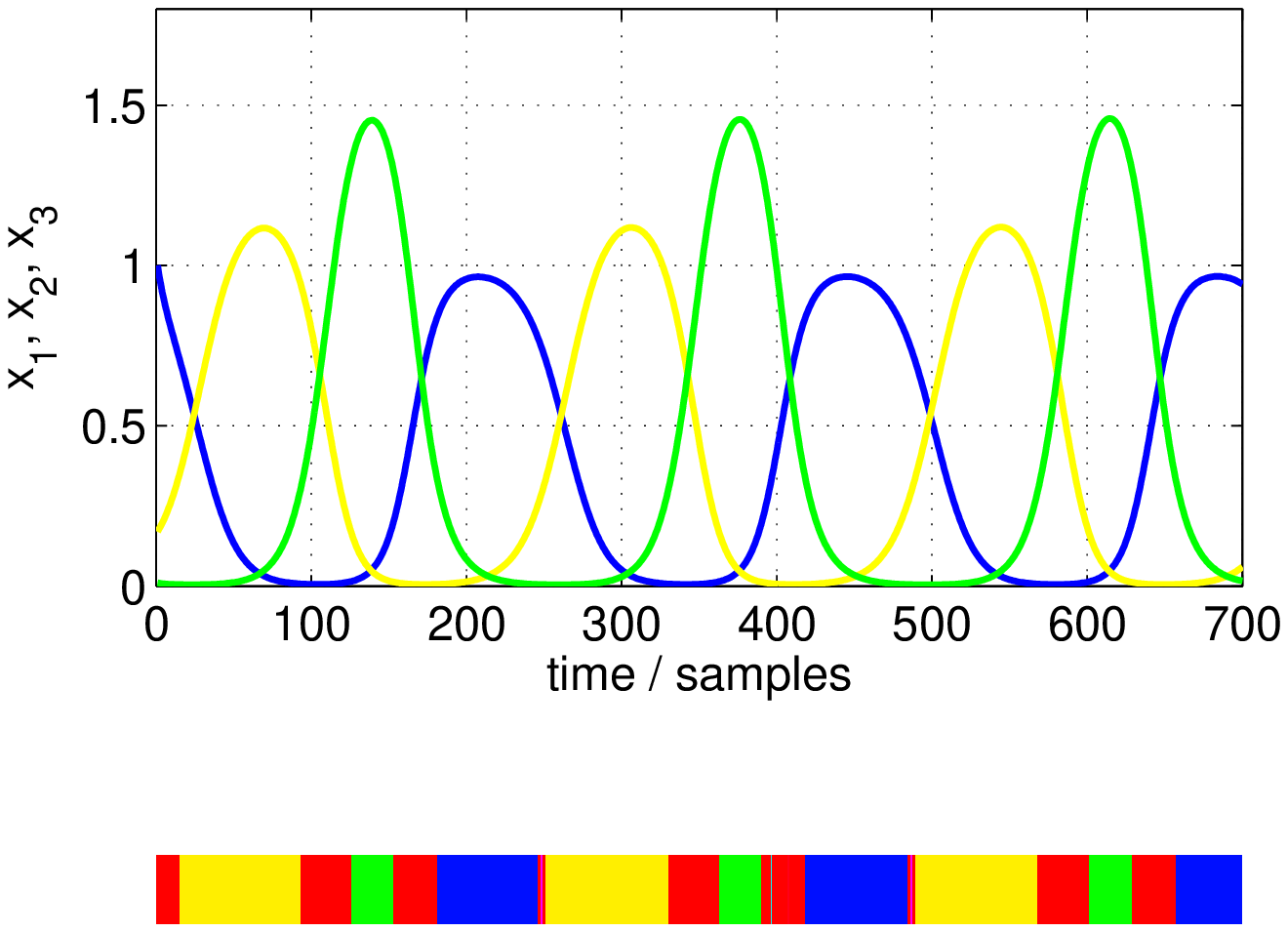}}
\caption{Stable heteroclinic contour. (a) Phase portrait with indicated recurrence domain partition. (b) Multivariate time series (upper panel) and symbolic segmentation into recurrence domains (bottom panel).}
\label{fig:shs}
\end{figure}

The heteroclinic contour in our simulation is governed by $\rho_{3,1} = \sigma_3 / \sigma_1 + 0.5$, $\rho_{2,1} = \sigma_2 / \sigma_1 - 0.5$ for saddle 1; $\rho_{1,2} = \sigma_1 / \sigma_2 + 0.5$, $\rho_{3,2} = \sigma_3 / \sigma_2 - 0.5$ for saddle 2; and $\rho_{2,3} = \sigma_2 / \sigma_3 + 0.5$, $\rho_{1,3} = \sigma_1 / \sigma_3 - 0.5$ for saddle 3. The system was initialized with $\vec{x} = (1, 0.17, 0.01)^T$ and numerically integrated over $[0, 100]$ with $\Delta t = 0.1429$ and the Runge-Kutta algorithm \texttt{ode45} of MATLAB.\textsuperscript{\texttrademark} Figure \ref{fig:shs} shows the resulting trajectory $\vec{x}(t)$ in  \Fig{fig:shs}(a) and the three coordinates, $x_1(t), x_2(t), x_3(t)$ in the upper panel of \Fig{fig:shs}(b).

Regarding the coordinates as ``recording channels'' of an (idealised) EEG experiment, we can identify the traces of $x_1$ (blue), $x_2$ (yellow), and $x_3$ (green) as ``ERP components'', peaking at the respective saddle nodes in phase space.

Applying our recurrence domain algorithm to the simulated data with $\varepsilon^* = 0.017$ yields the symbolic segmentation displayed in the bottom panel of \Fig{fig:shs}(b). Each peak is now assigned to one recurrence domain symbolised by the same colour. The corresponding phase space partition is depicted by the coloured balls in \Fig{fig:shs}(a).

One of the pertinent problems in recurrence analysis is optimising the ball size $\varepsilon$ \cite{Marwan11}. In our previous study \cite{GrabenHutt13}, we suggested to maximise the ratio of symbol entropy
\begin{equation}
\label{eq:entropy}
    H(\varepsilon) = - \sum_{k=1}^{M(\varepsilon)} p_k \log p_k
\end{equation}
over the cardinality of the symbolic repertoire $M(\varepsilon)$, where $p_k$ is the relative frequency of symbol $k$ in the symbolic sequence $s$. However, this utility function strongly penalises large alphabets through the denominator $M(\varepsilon)$ such that the proposed procedure is biased towards binary partitions which is not appropriate in most applications, such as our SHS example as well. Therefore, we here remedy this bias by an additional recoding of $s$. The symbolic dynamics obtained from recurrence grammar encoding still contains subsequences (or ``words'')  of monotonically increasing time indices, resulting from isolated $\varepsilon$-balls. These isolated balls, reflecting the transients of the dynamics, entail large symbol alphabets that are punished by the utility function above.

For a new optimisation method, we scan the sequence $s$ for monotonically increasing time indices, such as 4567 that are replaced by a new symbol 0 resulting in a subsequence 0000. Thereby, all isolated balls receive the same label 0, now indicating ``transient'' in the symbolic dynamics. This new symbol is plotted red in \Fig{fig:shs}; red segments and partition cells therefore denote transients in our encoding.

\begin{figure}[H]
\centering
\subfigure[]{\includegraphics[width=0.45\textwidth]{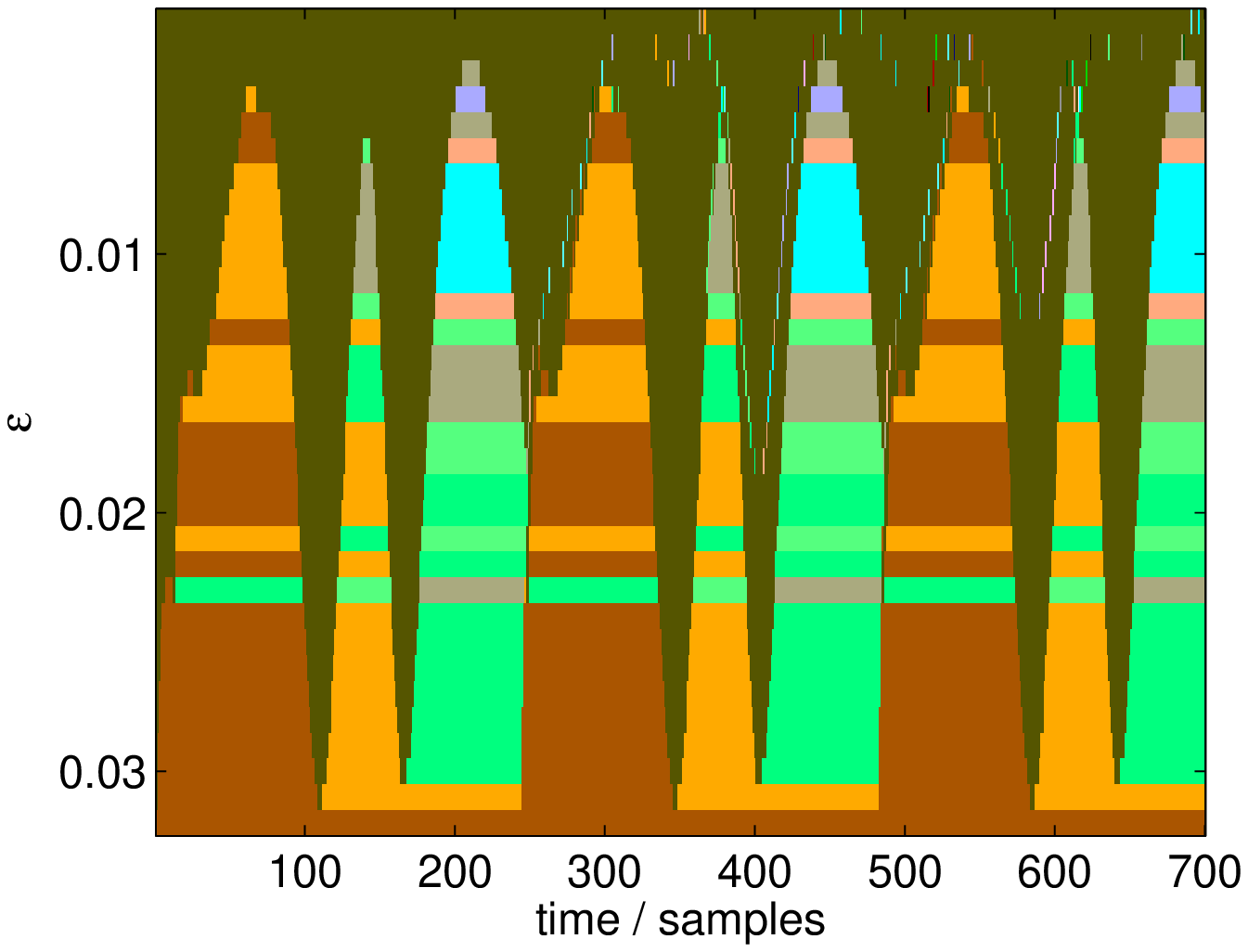}}
\subfigure[]{\includegraphics[width=0.45\textwidth]{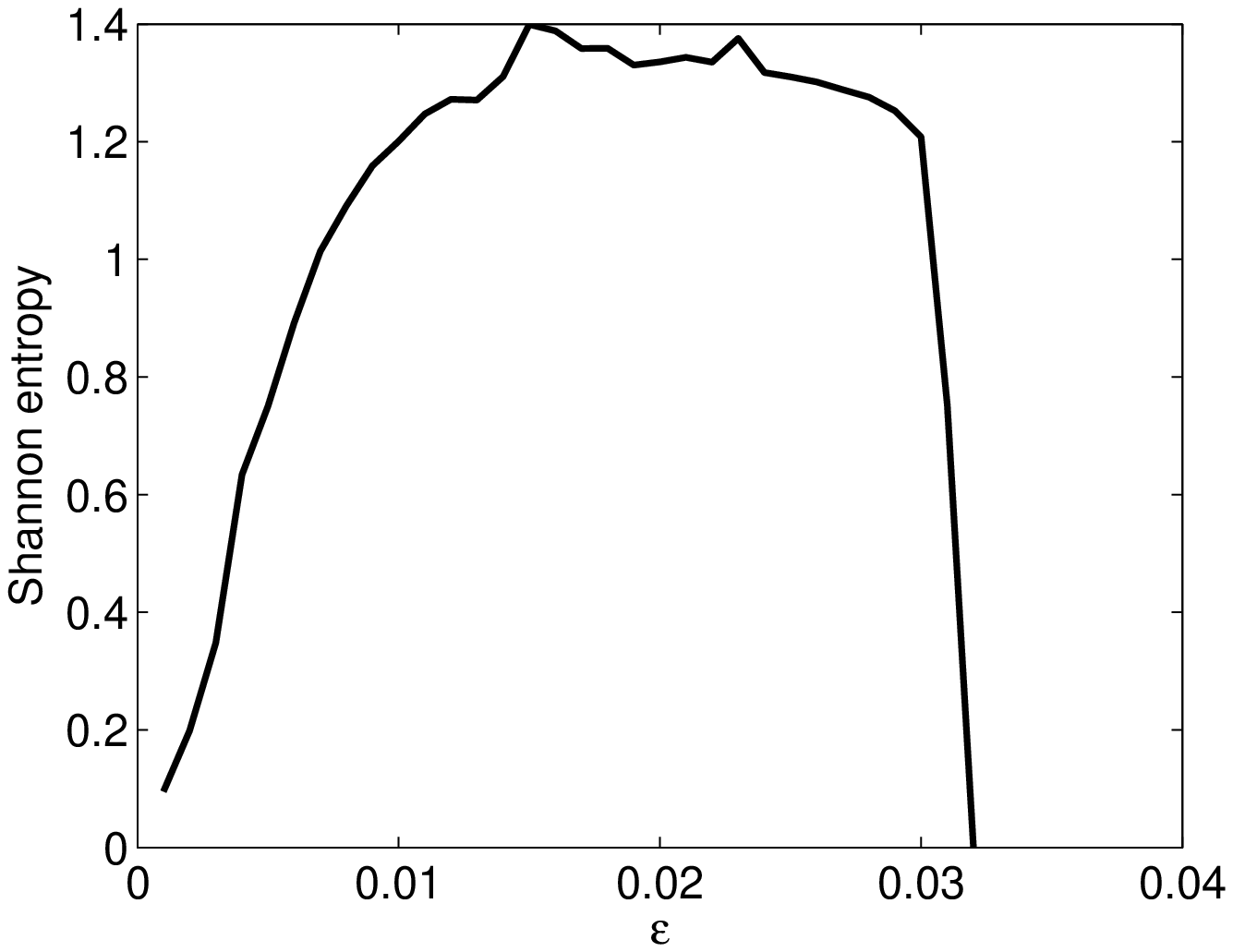}}
\caption{Dependency of recurrence domain partition from ball size $\varepsilon$. (a) Symbolic segmentation of time series from heteroclinic contour into recurrence domains for varying $\varepsilon$. Same colours encodes the same symbol. (b) Symbol entropy \pref{eq:entropy} dependent from $\varepsilon$.
}
\label{fig:epsopt}
\end{figure}

Hence, the optimisation algorithm performs a sequence of three encoding steps: (\emph{i}) rewriting by the recurrence grammar, (\emph{ii}) shifting the alphabet for closing gaps, (\emph{iii}) mapping transients onto 0. Figure \ref{fig:epsopt} shows the intermediate results of this optimisation procedure. In \Fig{fig:epsopt}(a) we plot the final symbolic sequences (called $s$, again) in dependence of the ball-size $\varepsilon$.

The symbolic dynamics plotted in \Fig{fig:epsopt}(a) starts with $\varepsilon = 0.001$ in the first row. Here, almost all balls remain isolated which is shown by the symbol 0 printed in olive here. For increasing $\varepsilon$, the number of recurrences increases due to the trajectory's slowing down in the vicinity of the saddle nodes. These saddles correspond to the ``tongues'' widening up to the recurrence domains. For some range of $\varepsilon$ the results are quite robust, indicating the stability of our algorithm. At the bottom row of \Fig{fig:epsopt}(a) $\varepsilon = 0.032$ where eventually all balls merge together into one domain covering the entire trajectory. This trivial domain is printed in dark brown here.

Computing the Shannon entropy \Eq{eq:entropy} for each row in \Fig{fig:epsopt}(a), yields the the distribution in \Fig{fig:epsopt}(b) which is relatively broad and exhibits several local maxima. Therefore, we suggest not to look for one of these (probably) spurious maxima for optimizing $\varepsilon$, but rather for the median of a symmetric distribution, or likewise for another quantile in case of an asymmetric entropy distribution. In this way we obtained the optimal ball size $\varepsilon^*$ for \Fig{fig:shs} through the median (the $q=0.5$ quantile).

\subsubsection{Hausdorff Partition Clustering}
\label{sec:hpc}

Lotka-Volterra systems as in \Eq{eq:lovolt} do not only possess stable heteroclinic sequences. These trajectories are additionally captured by stable heteroclinic channels (SHC) \cite{RabinovichEA08} that illustrate another important property of event-related brain potentials. Adopting our working hypothesis again, that ERP components correspond to saddle sets, or more generally, to recurrence domains in the brain's phase space \cite{HuttRiedel03, Hutt04}, we need to understand the emergence of ERP waveforms from ensemble averages over realisations of the system's dynamics starting from randomly distributed initial conditions \cite{Basar80, GrabenSaddyEA00}.

\begin{figure}[H]
\centering
\subfigure[]{\includegraphics[width=0.45\textwidth]{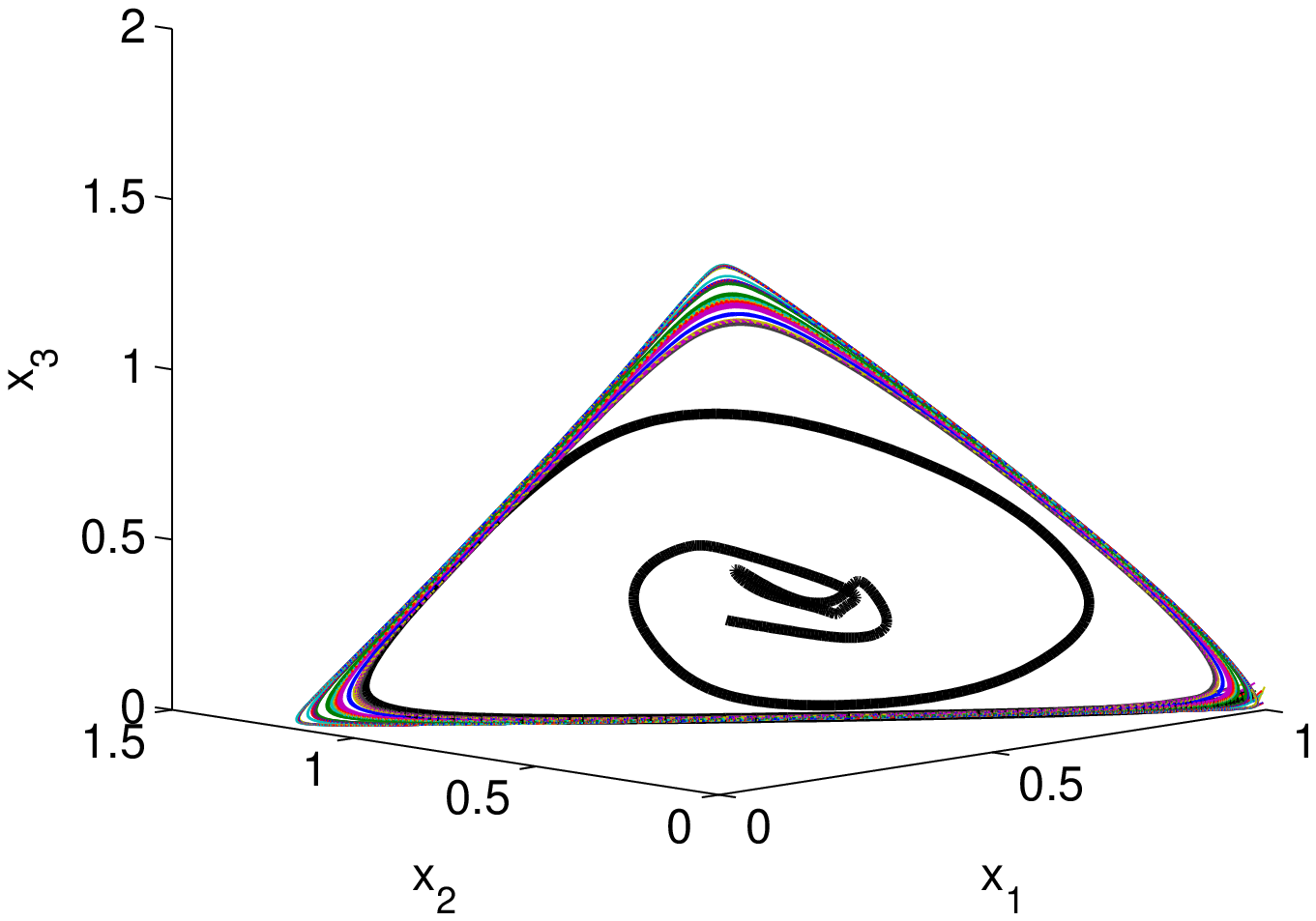}}
\subfigure[]{\includegraphics[width=0.45\textwidth]{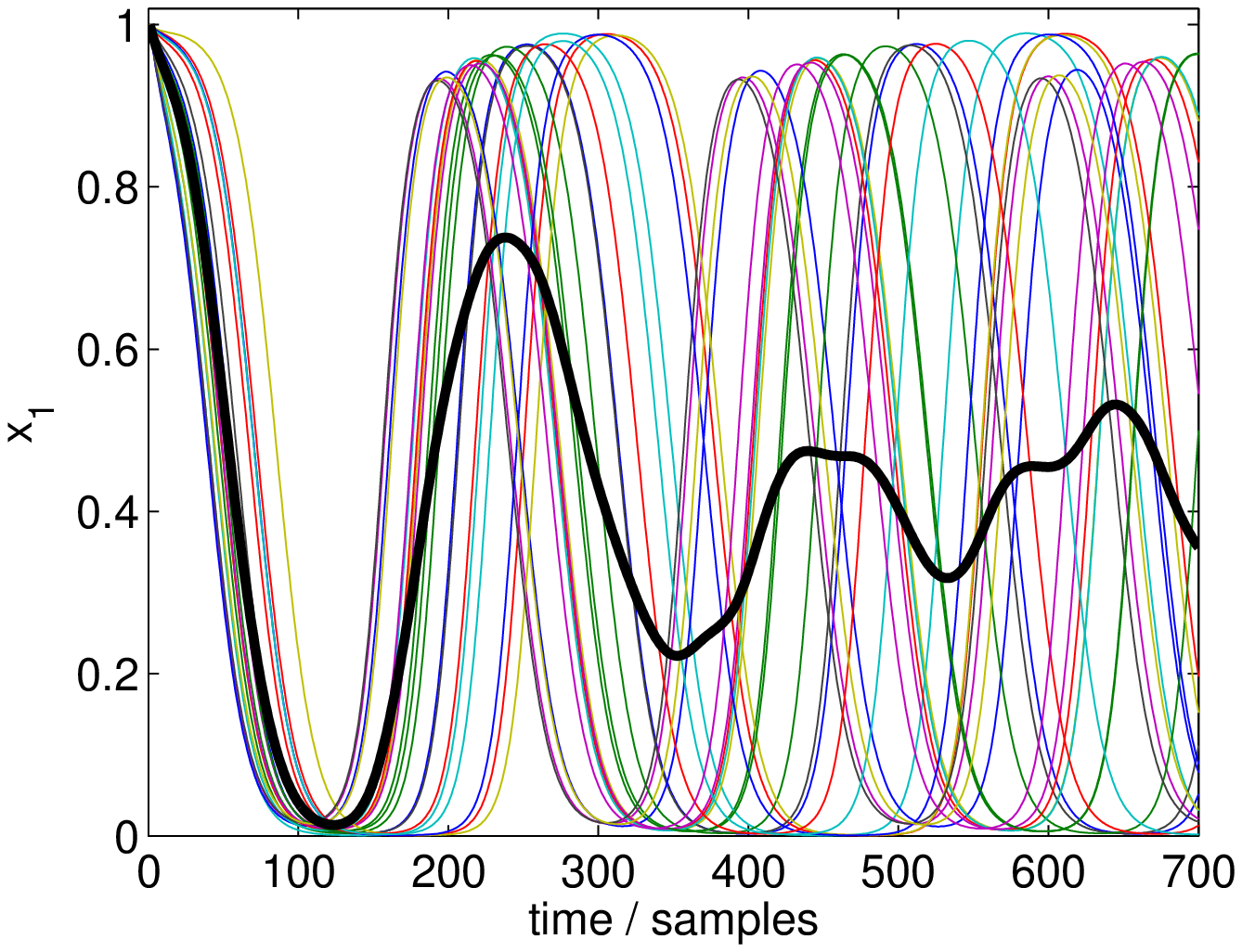}}
\caption{Stable heteroclinic channel (SHC). (a) Phase portrait of multiple realisations (coloured lines) and ensemble average (thick black line). (b) Projections onto $x_1$-axis.}
\label{fig:shc}
\end{figure}

To this end, we simulated an ensemble of 20 trajectories starting from randomly distributed initial conditions in the vicinity of the first saddle node (the blue recurrence domain in \Fig{fig:shs}). Figure \ref{fig:shc} shows the phase portrait of this ensemble (each realisation plotted in a different colour) in \Fig{fig:shc}(a) and the $x_3$-components in \Fig{fig:shc}(b). In addition we present the ensemble averages as the fat black curves in both pictures.

Figure \ref{fig:shc} nicely demonstrates that all simulated trajectories are confined to a  ``tube'', namely the SHC, stretched across the saddle nodes. This channel is stable against small perturbations in the initial conditions and also against additive noise \cite{RabinovichEA08}. However, the figure also shows the dispersion of the ensemble average which is inevitable in standard ERP analyses (yet see \cite{GrabenSaddyEA00, OuyangHerzmannEA11} for alternative approaches). This dispersion is due to the velocity differences dependent on the distance of the actual state from the saddles. The closer a state comes to the saddle the larger is its acceleration along the stable and its deceleration along the unstable manifold. Hence, the velocities with which states explore their available phase space crucially depend on the initial conditions. As the system evolves, the dynamics accumulates phase differences, deteriorating the averaged ensemble trajectory. This finding is consistent with the phase-resetting hypothesis of ERPs \cite{MakeigEA02} because phase-resetting takes place in the preparation of initial conditions.

Yet, our example is also relevant for the larger scope of nonlinear data analysis problems. Computing recurrence plots and subsequent recurrence quantification analysis (RQA) \cite{ZbilutZhihong00, Marwan11} is numerically rather expensive and therefore restricted to relatively short time series. In practice, a long-lasting time series is therefore cut into many segments for which recurrence analyses is carried out individually. The pertinent problem is then comparison and alignment of the individual results.

\begin{figure}[H]
\centering
\subfigure[]{\includegraphics[width=0.45\textwidth]{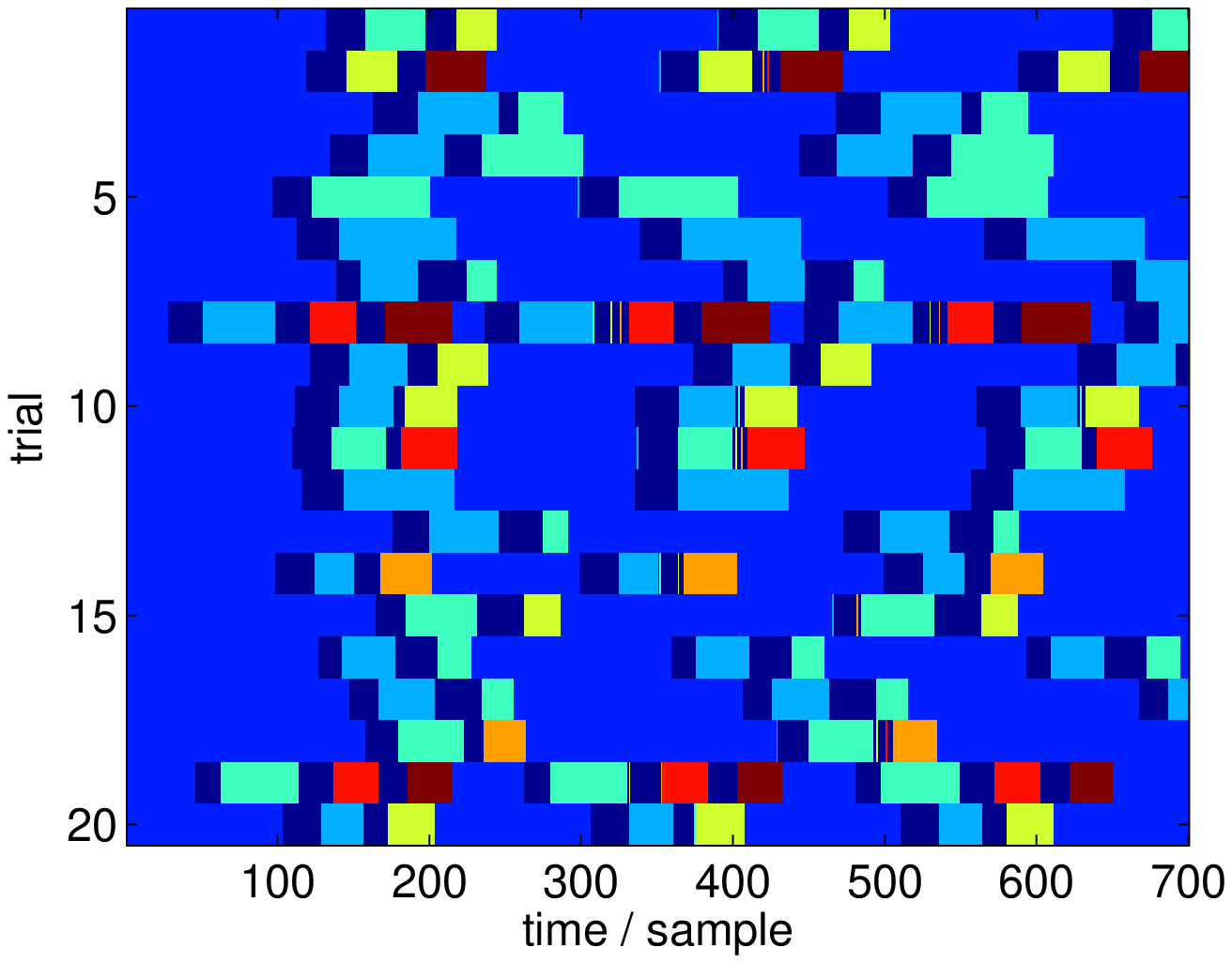}}
\subfigure[]{\includegraphics[width=0.45\textwidth]{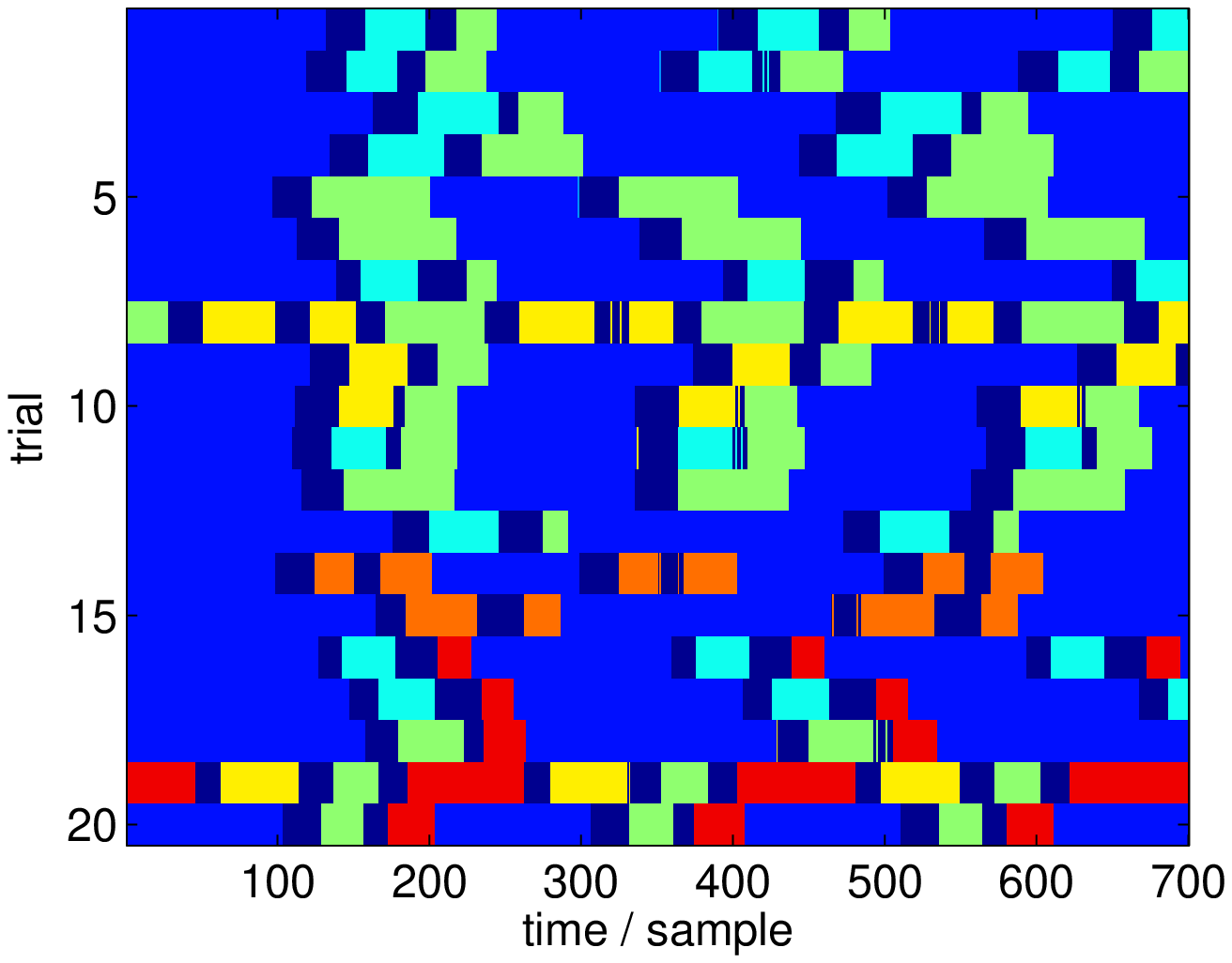}}
\caption{Symbolic segmentation of the realisations from \Fig{fig:shc} into recurrence domains. (a) before --- (b) after Hausdorff partition clustering. Dark Blue encodes transients.
}
\label{fig:hdcluster}
\end{figure}

In the framework of our symbolic recurrence domain analysis, we suggest here a solution based on a clustering algorithm whose results are presented in \Fig{fig:hdcluster}. In \Fig{fig:hdcluster}(a) we plot the symbolic sequences of the 20 SHS realisations from \Fig{fig:shc}, now encoded with the $q = 0.7$ quantile.

Figure \ref{fig:hdcluster}(a) shows different realisations using essentially the same colour palette which is due to the recurrence grammar and recoding algorithm. However, the same symbol (i.e. colour) does not necessarily refer to the same recurrence domain in the system's phase space (with one exception: transients are always encoded as 0, here plotted in dark blue). Therefore, we have to render the sequences in \ref{fig:hdcluster}(a) comparable with respect to corresponding regions in phase space.

For that aim, we first gather all discrete sampling points belonging to the same phase space cell $A_k$ that is labeled by the symbol $k$, i.e. $A_k = \{\vec{x}_i | s_i = k \} \subset X$. Since high-dimensional phase spaces are usually very sparsely populated with sampling points from experimental or simulated data, we then project the cells onto the $d$-dimensional hypersphere through $\vec{y}_i = \vec{x}_i / || \vec{x}_i ||$. This is also justified by the goal to clustering recurrent topographies of ERP data \cite{LehmannOzakiPal87, WackermannEA93, Pascual-MarquiMichelLehmann95}.

In the next step, the recurrence domain partitions $\mathcal{P}_1, \mathcal{P}_2$ of two subsequent realisations are merged together into a set system $\mathcal{Q} = \mathcal{P}_1 \cup \mathcal{P}_2$ by regarding all symbols (apart from 0) as being different. This system is in general not a partition any more. From the members of $\mathcal{Q}$ we then calculate the pairwise Hausdorff distances\footnote{
    Note that Wackermann et al. \cite{WackermannEA93} used a very similar method for clustering brain microstate topographies.
}
\begin{equation}
\label{eq:hddist}
    D_{ij} =  \max \{\max\{d(\vec{y}, A_j) ~|~ \vec{y} \in A_i \} , \max\{d(\vec{y}, A_i) ~|~ \vec{y} \in A_j \} \}
\end{equation}
where
\begin{equation}
\label{eq:hddist2}
    d(\vec{x}, A) =\min \{|| \vec{x} - \vec{y} || ~|~ \vec{y} \in A\}
\end{equation}
measures the ``distance'' of the point $\vec{x}$ from the compact set $A \subset X$. The Hausdorff distance between two compact sets vanishes, when they are intersecting. The recurrence domains taken into account here are clearly compact sets as they are finite unions of intersecting $\varepsilon$-balls.

From the pairwise Hausdorff distances \pref{eq:hddist} we compute the $\theta$-similarity matrix $\vec{S}$ with elements
\begin{equation}
\label{eq:sim}
    S_{ij} = \begin{cases}
    1 & \text{ if  }  D_{ij} < \theta \\
    0 & \text{ else }
    \end{cases}
\end{equation}
which has essentially the same properties as the recurrence matrix $\vec{R}$ from \Eq{eq:rp}. Therefore, we merge the members of $\mathcal{Q}$ into new partition cells by interpreting $\vec{S}$ as another recurrence grammar for rewriting large indices of $A_i$ by smaller ones from $A_j$ ($i>j$) when they are $\theta$-similar ($S_{ij} = 1$).

For numerical implementation we recursively decompose the ensemble of realisations into halves until only one or two trajectories are obtained. In the latter case, their recurrence domain partitions are merged together and subjected to the recurrence grammar given by the Hausdorff similarity matrix $\vec{S}$. The resulting partition contains the unions of similar recurrence domains with are passed to the next iteration of the algorithm. Although the algorithm shows nice convergence it is numerically very demanding using the MATLAB\textsuperscript{\texttrademark} interpreter language.

The result of the Hausdorff partition clustering algorithm applied to our SHC simulation is shown in \Fig{fig:hdcluster}(b) for $\theta = 0.7$. Now all symbols (colours) refer to the same recurrence domains in phase space (respectively to their topographic projections). Realisations \#1 -- \#12 exhibit very similar behaviours as they explore the same recurrence domains. Note also the differences in phase and duration in the individual realisations. Considered as a model of ERP data, \Fig{fig:hdcluster}(b) shows great resemblance to statically encoded ERPs \cite{GrabenSaddyEA00} where ERP components correspond to meandering vertical stripes.

\subsection{Event-related Brain Potentials}
\label{sec:erps}

The analysis of electroencephalographic data by means of symbolic dynamics techniques has a longstanding tradition. It could be traced back to Lehmann \cite{Lehmann71} and Callaway et al. \cite{CallawayHalliday73} who encoded extremal EEG voltages in a binary fashion and estimated the relative frequencies of these symbols across trials, respectively. The former approach led later to the so-called half-wave encoding \cite{GrabenFrisch04} while the latter was utilised through cylinder measures and word statistics \cite{GrabenSaddyEA00}. Further developments in this field were the symbolic resonance analysis \cite{GrabenKurths03, GrabenFrischEA05} and order pattern analyses \cite{BandtPompe02, KellerWittfeld04, SchinkelMarwanKurths07}. Also the segmentation of the EEG into quasistable ``brain microstates'' can be subsumed to these techniques \cite{LehmannOzakiPal87, WackermannEA93, Pascual-MarquiMichelLehmann95} which were recently combined with spectral clustering methods on approximate Markov chains \cite{AllefeldAtmanspacherWackermann09, Froyland05, GaveauSchulman06}.

In this section we reanalyse an ERP experiment on the processing of ungrammaticalities in German \cite{FrischHahneFriederici04} (see \cite{FrischGrabenSchlesewsky04, FrischGraben05, DrenhausGrabenEA06, GrabenGerthVasishth08} for other studies on symbolic dynamics of language-related brain potentials). Frisch et al. \cite{FrischHahneFriederici04} examined processing differences for different violations of lexical and grammatical rules. Here we focus only on the contrast between a so-called \emph{phrase structure violation} \ref{ex:psv}, indicated by the asterisk, in comparison to grammatical control sentences \ref{ex:cor}.

\setcounter{ExNo}{\value{equation}}

\ex.
    \ag. Im Garten wurde oft \textbf{gearbeitet} und \dots \\
    In{ }the garden was often \textbf{worked} and \dots \\
    ``Work was often going on in the garden \dots''
    \label{ex:cor}
    \bg.  * Im Garten wurde am \textbf{gearbeitet} und \dots \\
    In{ }the garden was on-the \textbf{worked} and \dots \\
    ``Work was on-the going on in the garden \dots''
    \label{ex:psv}

Sentences of type \ref{ex:psv} are ungrammatical in German because the preposition \emph{am} is followed by a 	past participle instead of a noun. A correct continuation could be, e.g., \emph{im Garten wurde am \textbf{Zaun} gearbeitet} (``work was going on in the garden at the fence'') with \emph{am Zaun} (``at the fence'') comprising an admissible prepositional phrase.

The ERP study was conducted with 17 subjects in a visual word-by-word presentation paradigm. Subjects were presented with 40 structurally identical examples per condition. The critical word in all conditions was the past participle printed in bold font in the above. EEG and additional electroocologram (EOG) for controling eye-movement were recorded with a 64 electrode montage from which 59 EEG electrodes were selected for recurrence domain analysis.

\subsubsection{Grand Averages}
\label{sec:grav}

\begin{figure}[H]
\centering
\includegraphics[width=\textwidth]{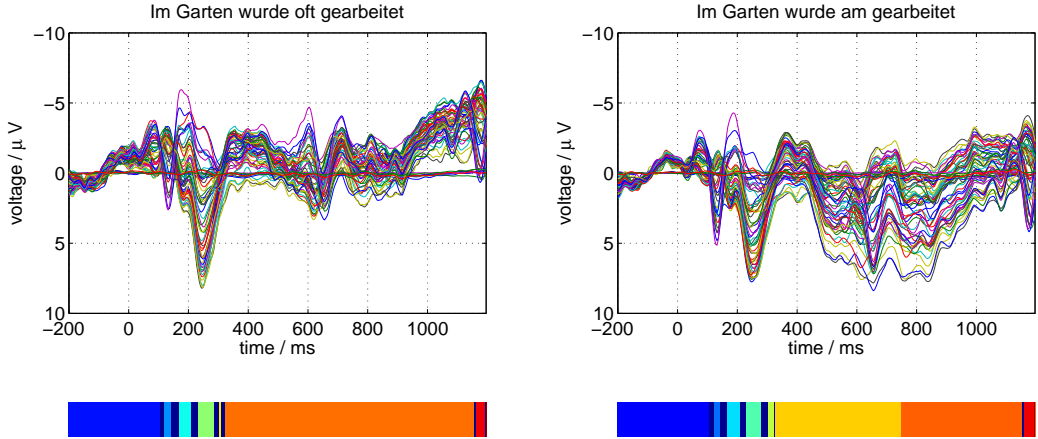}
\caption{ERP grand averages and recurrence domain segmentation from \cite{FrischHahneFriederici04}. (Left panels) Correct condition \ref{ex:cor}. (Right panels) Phrase structure violation condition \ref{ex:psv}. (Upper panels) grand average ERPs over 16 subjects. Each trace showing one recording channel. (Bottom panels) Symbolic dynamics of recurrence domains.}
\label{fig:erps}
\end{figure}

First we carry out the recurrence domain analysis for the grand averages over all 17 subjects which are presented in the upper panels of \Fig{fig:erps}. At the left-hand side the multivariate ERP time series for the correct condition \ref{ex:cor} are plotted as coloured traces. The same for the time series of the phrase structure violation condition \ref{ex:psv} at the right-hand side. Both plots start 200 ms before stimulus presentation at $t = 0$. In both conditions similar N100/P200 ERP complexes are evoked. Yet  condition \ref{ex:psv} exhibits a large positive P600 ERP at central and posterior electrodes around 600 ms after stimulus presentation.

For the recurrence domain analysis we regard the observation space \cite{BirkhoffNeumann36} spanned by all EEG-channel voltages $\vec{v}$ as the system's phase space \cite{Stam05} and compute entropy distributions $H(\varepsilon)$ \pref{eq:entropy} in the range $\varepsilon \in [0.2 \mu\mathrm{V}, 5\mu\mathrm{V}]$ which are very asymmetric. Therefore we chose the $q = 0.77$ quantile for optimisation by visual inspection of both segmentations, yielding $\varepsilon^* = 2.2\mu\mathrm{V}$ for the control condition. We then compute the recurrence grammars and the resulting segmentation into recurrence domains with this parameter for both conditions. The results are shown in the bottom panels of \Fig{fig:erps}.

From \Fig{fig:erps}(bottom) we draw three important conclusions: (\emph{i}) the prestimulus interval and also the first 100 ms are assigned to the same recurrence domain, reflecting the resting state brain activity that is not related to the stimulus processing. (\emph{ii}) In the time window from 100 ms to 300 ms a similar partitioning into successive recurrence domains is observed for both conditions. This indicates similarities in the early attentional processes assigned to the N100/P200 ERP complexes \cite{Woods90}. (\emph{iii}) Both conditions differ crucially after 300 ms. In the correct control condition \ref{ex:cor} there is only one large recurrence domain in this time window, whereas the phrase structure violation condition \ref{ex:psv} exhibits two  recurrence domains: a first one in the time window of lexical access and syntactic diagnoses processes around 400 -- 700 ms, and another one in the time window of syntactic repair and reanalysis processes around 700 -- 1200 ms which is consistent with other findings about sub-components of the late positivity \cite{DrenhausGrabenEA06}.

\subsubsection{Single Subject Clustering}
\label{sec:sisu}

\begin{figure}[H]
\centering
\subfigure[]{\includegraphics[width=0.45\textwidth]{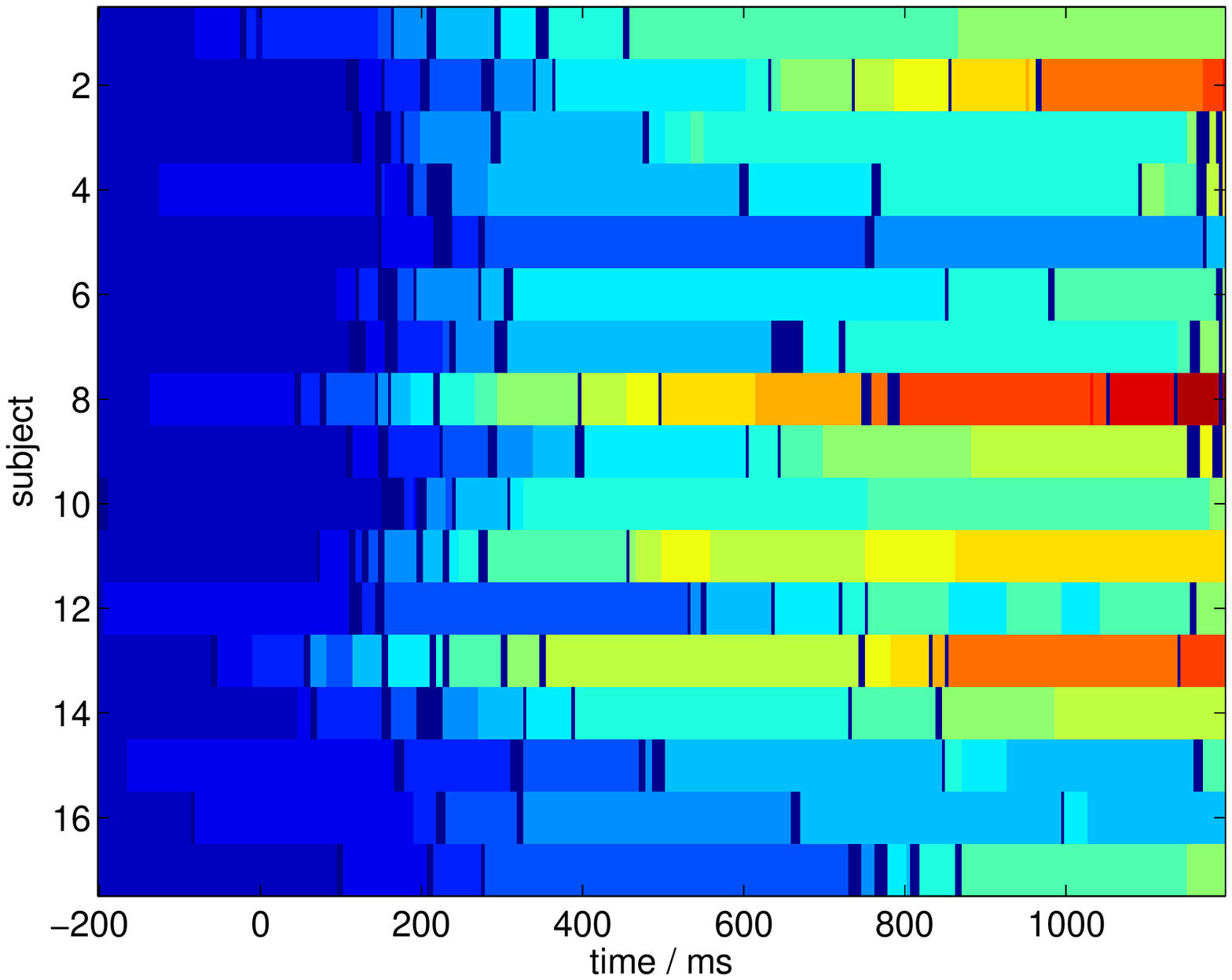}}
\subfigure[]{\includegraphics[width=0.45\textwidth]{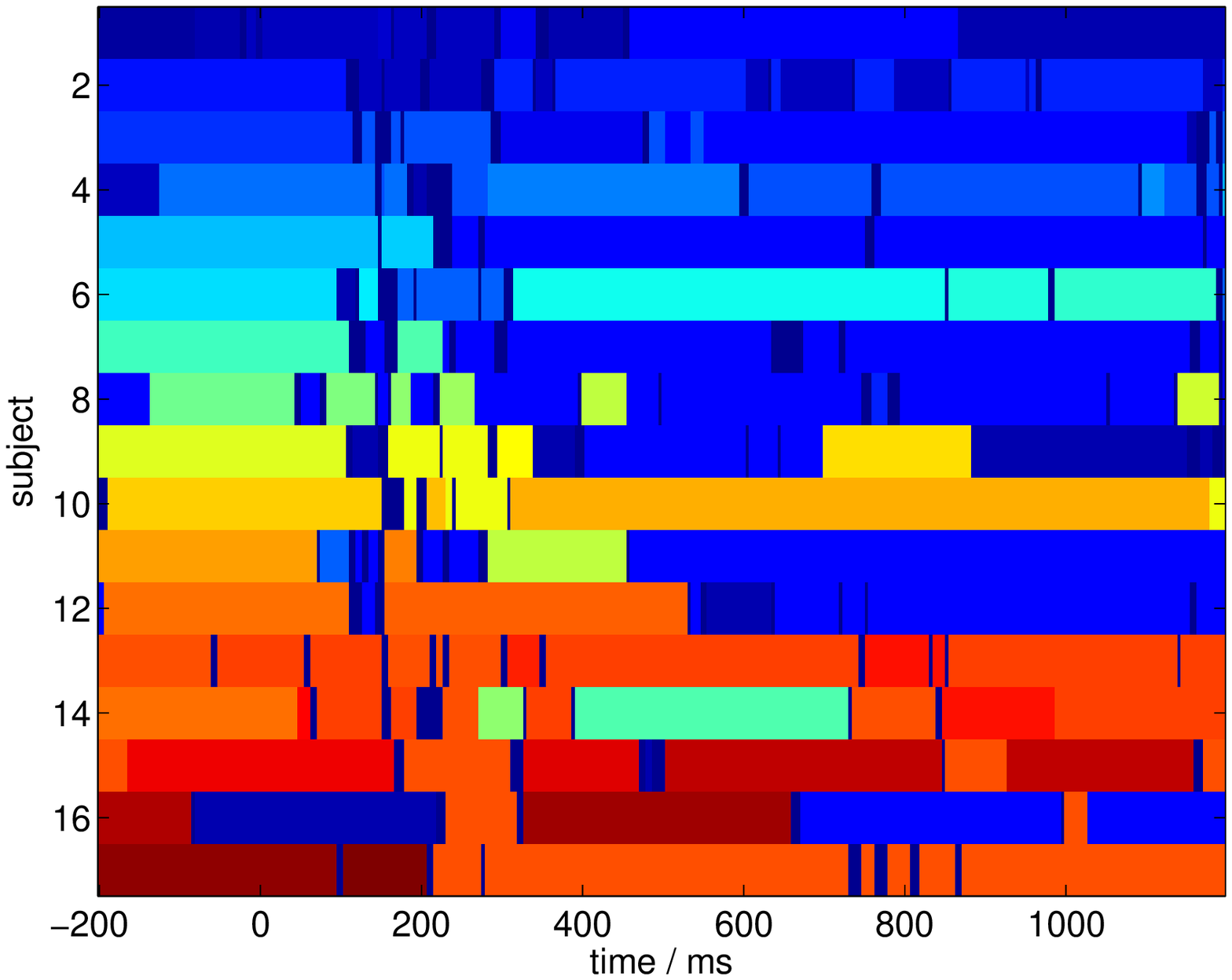}}
\caption{Hausdorff partition clustering of single subject ERPs for condition \ref{ex:psv}. (a) Symbolic dynamics of individual recurrence domains of 17 subject ERPs. (b) Recurrence domain symbolic dynamics after Hausdorff partition clustering.}
\label{fig:sisuhd}
\end{figure}

Finally, we carry out the Hausdorff partition cluster analysis for the phrase structure violation condition \ref{ex:psv}. Here, we present only a proof-of-concept using the single subject ERPs. All individual single subject ERPs of condition \ref{ex:psv} are subjected to the same recurrence analysis with the $q = 0.77$ quantile of the entropy distribution \pref{eq:entropy}. The results are shown in \Fig{fig:sisuhd}(a).

Figure \ref{fig:sisuhd}(a) displays large differences between individual subject ERPs although some resemblance with the grand average recurrence domains from \Fig{fig:erps}(bottom-right) are present, e.g. in subjects \#1, \#5, \#13, and \#17, exhibiting the expected segmentation. As every symbol must be treated differently across realisations, the cardinality of the alphabet for this representation is about 272.

In order to obtain a good clustering despite of the apparent differences, we chose a similarity threshold $\theta = 0.985$ for our Hausdorff algorithm. The resulting clustering of recurrence domain topographies is depicted in \ref{fig:sisuhd}(b). The cardinality of the alphabet is drastically reduced to 54 different symbols. Obviously, some recurrence domains are shared by many realisations, as reflected by the repeating symbol appearing in medium blue in subjects \#1 -- \#3,  \#7 -- \#9, as well as \#11, \#12, and \#16 around 400 -- 800 ms. However, the desired ``meandering vertical stripes'' are not yet visible.

For further evaluation, the symbolic dynamics obtained from different experimental conditions, different subjects and eventually from single trials could be subjected to statistical analysis by computing word distributions, cylinder entropies and statistical hypothesis tests (e.g. $\chi^2$ tests on word distributions or permutation tests on symbolic ensembles) \cite{GrabenSaddyEA00, GrabenFrischEA05}. We leave this as well as further parameter search for optimising encodings for a future publication.

\section{Conclusion}
\label{sec:conc}

Starting from the working hypothesis that event-related brain potentials (ERP) reflect quasistationary states in brain dynamics, we have elaborated an earlier proposal for the detection of quasistationary states and for the segmentation of electroencephalographic time series into recurrence domains. This segmentation technique interprets the recurrence matrix as a rewriting grammar that applies to the time indices of an ERP data set. After further recoding steps, the quasistationary states are detected as recurrence domains in phase space as indicated by a symbolic dynamics. Moreover, we have suggested a method for the alignment and unification of multiple EEG trials (i.e. single subject ERPs) by means of Hausdorff partition clustering.

We think that the presented methods could be of significance for the greater community of nonlinear time series analysis, as we have addressed some pertinent problems in recurrence analysis and symbolic dynamics. We have suggested an optimisation procedure for choosing the ball size of recurrence plots by maximising the entropy of the symbol distribution that is obtained by applying recurrence grammars and subsequent recoding of transients. Also the alignment and comparison of recurrence plots of different time windows or different realisations was an unsolved problem so far. We solved this problem by merging together recurrence domains from different time series upon their similarity with respect to the Hausdorff distance either in phase space, or, in projection onto the unit sphere.

\section*{Acknowledgments}
We thank the guest editors for their kind invitation to contribute to this issue of the \emph{Philosophical Transactions}. In this study we reanalysed a language processing EEG data set by courtesy of Stefan Frisch, Anja Hahne and Angela Friederici. AH and PbG acknowledge funding from the European Research Council for support under the European Union's Seventh Framework Programme (FP7/2007-2013) ERC grant agreement No.257253. In addition, PbG acknowledges support by a Heisenberg Fellowship of the German Research Foundation DFG (GR 3711/1-2).


\end{document}